\documentclass[a4paper,UKenglish]{lipics}
\usepackage{mylogicv03,graphicx,hyperref}

\def\nto{\mathchar `\-}
\standardlogic

\title{Teaching natural deduction in the right order with Natural Deduction Planner}

\author[1]{Jeremy Seligman}
\author[2]{Declan Thompson}
\affil[1]{The University of Auckland\\
	New Zealand\\
	\texttt{j.seligman@auckland.ac.nz}}
\affil[2]{The University of Auckland\\
	New Zealand\\
	\texttt{dtho139@aucklanduni.ac.nz}}
\authorrunning{J. Seligman and D. Thompson} 

\Copyright{Jeremy Seligman and Declan Thompson}


\serieslogo{logo_ttl}
\volumeinfo
  {M. Antonia {Huertas}, Jo\~ao {Marcos}, Mar\'ia {Manzano}, Sophie
{Pinchinat}, \\
  Fran\c{c}ois {Schwarzentruber}}
  {5}
  {4th International Conference on Tools for Teaching Logic}
  {1}
  {1}
  {199}
\EventShortName{TTL2015}

\begin{document}

 \maketitle
 
\begin{abstract}
\noindent
We describe a strategy-based approach to teaching natural deduction using a notation that emphasises the order in which deductions are constructed, together with a {\LaTeX} package and Java app to aid in the production of teaching resources and classroom demonstrations. Our approach is aimed at students with little exposure to mathematical method and has been developed while teaching undergraduate classes for philosophy students over the last ten years. 
\end{abstract}

\section{Natural Deduction as a Creative Process}
 
Teaching modern logic to students with little background in mathematics is notoriously hard. The philosophy student, adept at reading complex prose and composing artful essays is usually not well prepared for manipulating symbols and constructing rigorous proofs of theorems.  Acquisition of at least the following three skills are needed. 

The first is using the language of propositional and predicate logic to represent one's thoughts in formal notation and understand what has been written by others. This is usually achieved by learning to translate to and from natural language. Many resources are available. 

The second is manipulating the symbols of formal notation according to precise rules. This is a basic skill necessary for almost all of logic, from applying mechanical methods of argument evaluation to acquiring an appreciation of the autonomy of the syntactic realm, without which the major theoretical results of logical theory cannot be understood. It can be acquired by learning how to produce truth tables, truth trees, and in many other ways. Again many resources are available. 

The third is reading and writing rigorous arguments, of the kind used in mathematics.  This is a much more difficult skill to acquire, requiring mastery of the first two skills, and in addition, a level of mathematical maturity that is attained by mathematics students only after years of practice with algebra, geometry, analysis, etc. Consequently, this side of logic education is often neglected by philosophy undergraduate programmes. Although many introductory textbooks include some discussion of logical theory, such as soundness and completeness, the emphasis is on understanding the theorems rather than developing the skills to prove them.  Few are aimed directly at acquiring the skill of creating proofs from scratch.\footnote{A notable exception is ``How to Prove It: A Structured Approach'' \cite{velleman1994prove,velleman2006prove}.  There are also countless introductions to mathematical method, e.g. ``The Nuts and Bolts of Mathematical Proof" \cite{cupillari2005nuts}, ``How to Read and Do Proofs: An Introduction to Mathematical Thought Processes'' \cite{solow2002read}.}

One solution is to require logic students to take a substantial number of courses in mathematics, so that they acquire the necessary skills in the same way as mathematics students. In the long run, a broad experience with mathematical methods is certainly useful for research in logic, if not absolutely essential. But the huge gap that must be filled is daunting and dispiriting for most philosophy students, most of whom decide that it is just too big to breach. 

Is there another solution?  The obvious candidate is to teach students the skill of rigorous argumentation using the very formalisms that they have already learned: propositional and predicate logic. From a theoretical perspective, we know that our various systems of deduction can duplicate all that a mathematics student learns by a much more indirect and less explicit route. Why then is it so difficult for a philosophy student who has learned a formal system of deduction to transfer her skills to the production of informally rigorous arguments of the kind needed for progress in her subject?  

It is generally recognised that axiomatic systems, while elegant and theoretically parsimonious, are wholly inappropriate for learning deduction.  Instead, most logic programmes for philosophy students include some system of natural deduction, in which axioms are replaced by rules which mirror patterns of reasoning used in natural language argumentation. In the classic approach of Irving Copi, numerous rules are added, so as capture as many such patterns as possible.\footnote{``Introduction to Logic" \cite{copi2013introduction} is now in its 14th edition. Many other textbooks on natural deduction employ a similarly lengthy list of rules.}
Yet there is often an insufficient level of attention to any systematic discussion of the process of \emph{creating} deductions.  Typically, students are given an introduction to the rules, motivated by their natural language correlates, a few examples of complete deductions, and are then left to fend for themselves on a large number of exercises, with the hope that they will develop their own strategies by trial and error. 

An alternative is to teach the \emph{strategies} of creation explicitly.  As well as helping students to learn formal deduction, these are the strategies that will prepare the student for the harder task of creating rigorous informal arguments of the kind needed to do postgraduate work in logic, and which mathematics students learn implicitly through their application to a wide range of mathematical topics. Teachers of natural deduction in the traditional style may be fully aware of this point, but the effective learning of explicit strategy is made almost impossibly hard by several factors. 

The first is simply the number of rules used by logic textbooks aimed at mirroring patterns in informal reasoning, which include both proof by cases (Disjunction Elimination, \rulename{\disop E}), and Disjunctive Syllogism, if not also Constructive Dilemma.  While each of these is relatively easy to explain in isolation, the more rules, the harder it is to master their strategic interactions, which the student must consider when creating her own deductions.  

A second, related factor is the lack of structure to the set of rules. From the perspective of teaching strategy, one would prefer a simpler set of rules, organised in a way that corresponds to patterns of use in the creation of deductions, and exactly this is provided by Gentzen's original system, which uses the idea of introduction and elimination to expose the structure and symmetry of proof.  There is no room here to explain and justify this choice, so we will just mention a few relevant points. Firstly, the fact that the intuitionistic fragment  of the system has only a pair of rules for each logical operator allows one to develop general strategies: one for Introduction rules and one for Elimination rules, concerning the management of resources and simplification of goals. Moreover, an orthogonal classification of rules allows us to distinguish between cases in which a choice is required (e.g., \rulename{\disop I} and \rulename{\qeop I}) and those that are `automatic', in the sense that they can be applied without the need for further choice. Even the symmetry-breaking oddity of the non-intuitionistic rule \rulename{\negop\negop E}, which can be applied to any conclusion, raises an important strategic question: how to manage the creative steps of deduction? And this leads to an explicit discussion of back-tracking in problem solving and the need to recognise dead-ends. While such matters of strategy are implicit in more complicated systems, they are highlighted in systems in which the number of rules is small and well-balanced. 

Standard presentations of Gentzen-style natural deduction, such as those of Fitch or Lemon, still have an important deficiency. Designed for reading rather than writing, the argument is displayed with the premises at the top, the conclusion at the bottom and with each line justified by lines higher up on the page, according to a formal rule. This makes the process of checking the correctness of the deduction relatively easy, but the process of generating the deduction itself unnecessarily hard. 
\begin{center}
\begin{tabular}{lp{6cm}}
$\begin{array}[t]{@{}l@{}l@{}lll}
          &           &  {1.}  &  {\dis{\neg{p}}{\neg{q}}} & {\rulename{Prem}}\\
\ndcorner & \ndhrule  &  {2.}  &  {\con{p}{q}}             & {\rulename{Ass}}\\
\ndvrule  & \ndcorner &  {3.}  &  {\neg{p}}                & {\rulename{Ass}}\\
\ndvrule  & \ndvrule  &  {4.}  &  {p}                      & {2,\rulename{\conop E}}\\
\ndvrule  & \ndvrule  &  {5.}  &  {\falsum}                & {3,4,\rulename{\negop E}}\\
\cline{2-4}
\ndvrule  & \ndcorner &  {6.}  &  {\neg{q}}                & {\rulename{Ass}}\\
\ndvrule  & \ndvrule  &  {7.} &  {q}                      & {2,\rulename{\conop E}}\\
\ndvrule  & \ndvrule  &  {8.}  &  {\falsum}                & {6,7,\rulename{\negop E}}\\
\cline{2-4}
\ndvrule  &           &  {9.}  &  {\falsum}                & {1,3\nto 5,6\nto 8,\rulename{\disop E}}\\
\cline{1-4}
          &           &  {10.}  &  {\neg{\con{p}{q}}}       & {2\nto 9,\rulename{\negop I}}\\
\end{array}$
&
\emph{On the left is a correct deduction using a version of Gentzen's rules. Hypothetical reasoning is indicated by marking the assumption (\rulename{Ass}) and a vertical bracket ending below the hypothetical conclusion.  The symbol $\falsum$ is used to mark a contradiction.}
\end{tabular}
\end{center}
Information about the process of \emph{creating} the deduction is lost in this representation, which wrongly suggests  that it was written from top to bottom, starting with the premises and ending in the conclusion.  (One of the most common mistakes made by students is to follow this order.) There is no record of the strategies used to construct the deduction; no record even of the order in which it was constructed. The student who fails to produce her own deduction of ${\neg{\con{p}{q}}}$ from $\dis{\neg{p}}{\neg{q}}$ will not learn much  from looking at the above solution. 

If we were to display a full sequence of steps leading to the creation of this deduction, we might write the following:
\begin{center}
\scalebox{0.7}{
\begin{tabular}{lclcl}
$\begin{array}[c]{@{}l@{}l@{}lll}
          &           &  {1.}  &  {\dis{\neg{p}}{\neg{q}}} & {\rulename{Prem}}\\
\\ \\ \\ \\ \\ \\ \\ \\
          &           &  {10.}  &  {\neg{\con{p}{q}}}       \\
\end{array}$
& ${\begin{array}{c}\rulename{\negop I}\\ \leadsto\end{array}}\quad$ &
$\begin{array}[c]{@{}l@{}l@{}lll}
          &           &  {1.}  &  {\dis{\neg{p}}{\neg{q}}} & {\rulename{Prem}}\\
\ndcorner & \ndhrule  &  {2.}  &  {\con{p}{q}}             & {\rulename{Ass}}\\
\ndvrule  &\\ 
\ndvrule  &\\ 
\ndvrule  &\\ 
\ndvrule  &\\ 
\ndvrule  &\\ 
\ndvrule  &\\ 
\ndvrule  &           &  {9.}  &  {\falsum}                \\
\cline{1-4}
          &           &  {10.}  &  {\neg{\con{p}{q}}}       & {2\nto 9,\rulename{\negop I}}\\
\end{array}$
& ${\begin{array}{c}\rulename{\disop E}\\ \leadsto\end{array}}\quad$ &
$\begin{array}[c]{@{}l@{}l@{}lll}
          &           &  {1.}  &  {\dis{\neg{p}}{\neg{q}}} & {\rulename{Prem}}\\
\ndcorner & \ndhrule  &  {2.}  &  {\con{p}{q}}             & {\rulename{Ass}}\\
\ndvrule  & \ndcorner &  {3.}  &  {\neg{p}}                & {\rulename{Ass}}\\
\ndvrule  & \ndvrule  &  \\
\ndvrule  & \ndvrule  &  {5.}  &  {\falsum}                \\
\cline{2-4}
\ndvrule  & \ndcorner &  {6.}  &  {\neg{q}}                & {\rulename{Ass}}\\
\ndvrule  & \ndvrule  & \\
\ndvrule  & \ndvrule  &  {8.}  &  {\falsum}                \\
\cline{2-4}
\ndvrule  &           &  {9.}  &  {\falsum}                & {1,3\nto 5,6\nto 8,\rulename{\disop E}}\\
\cline{1-4}
          &           &  {10.}  &  {\neg{\con{p}{q}}}       & {2\nto 9,\rulename{\negop I}}\\
\end{array}$
\end{tabular}}
\end{center}
\begin{center}
\scalebox{0.7}{
\begin{tabular}{clcl}
${\begin{array}{c}\rulename{\negop E},\rulename{\conop E}\\ \leadsto\end{array}}\quad$ &
$\begin{array}[c]{@{}l@{}l@{}lll}
          &           &  {1.}  &  {\dis{\neg{p}}{\neg{q}}} & {\rulename{Prem}}\\
\ndcorner & \ndhrule  &  {2.}  &  {\con{p}{q}}             & {\rulename{Ass}}\\
\ndvrule  & \ndcorner &  {3.}  &  {\neg{p}}                & {\rulename{Ass}}\\
\ndvrule  & \ndvrule  &  {4.}  &  {p}                      & {2,\rulename{\conop E}}\\
\ndvrule  & \ndvrule  &  {5.}  &  {\falsum}                & {3,4,\rulename{\negop E}}\\
\cline{2-4}
\ndvrule  & \ndcorner &  {6.}  &  {\neg{q}}                & {\rulename{Ass}}\\
\ndvrule  & \ndvrule  &  \\
\ndvrule  & \ndvrule  &  {8.}  &  {\falsum}                \\
\cline{2-4}
\ndvrule  &           &  {9.}  &  {\falsum}                & {1,3\nto 5,6\nto 8,\rulename{\disop E}}\\
\cline{1-4}
          &           &  {10.}  &  {\neg{\con{p}{q}}}       & {2\nto 9,\rulename{\negop I}}\\
\end{array}$
& ${\begin{array}{c}\rulename{\negop E},\rulename{\conop E}\\ \leadsto\end{array}}\quad$ &
$\begin{array}[c]{@{}l@{}l@{}lll}
          &           &  {1.}  &  {\dis{\neg{p}}{\neg{q}}} & {\rulename{Prem}}\\
\ndcorner & \ndhrule  &  {2.}  &  {\con{p}{q}}             & {\rulename{Ass}}\\
\ndvrule  & \ndcorner &  {3.}  &  {\neg{p}}                & {\rulename{Ass}}\\
\ndvrule  & \ndvrule  &  {4.}  &  {p}                      & {2,\rulename{\conop E}}\\
\ndvrule  & \ndvrule  &  {5.}  &  {\falsum}                & {3,4,\rulename{\negop E}}\\
\cline{2-4}
\ndvrule  & \ndcorner &  {6.}  &  {\neg{q}}                & {\rulename{Ass}}\\
\ndvrule  & \ndvrule  &  {7.} &  {q}                      & {2,\rulename{\conop E}}\\
\ndvrule  & \ndvrule  &  {8.}  &  {\falsum}                & {6,7,\rulename{\negop E}}\\
\cline{2-4}
\ndvrule  &           &  {9.}  &  {\falsum}                & {1,3\nto 5,6\nto 8,\rulename{\disop E}}\\
\cline{1-4}
          &           &  {10.}  &  {\neg{\con{p}{q}}}       & {2\nto 9,\rulename{\negop I}}\\
\end{array}$
\end{tabular}}
\end{center}
This is much too cumbersome for practical use in textbooks, and leaves the assignment of line numbers somewhat mysterious. How to know the deduction will use ten lines? But a simple change in notation can help. Instead of numbering the lines of the deduction from top to bottom, we number them in the order they were created.  The above sequence can then be represented with just one deduction, as shown:
$$\begin{array}[t]{@{}l@{}l@{}lll}
          &           &  {1.}  &  {\dis{\neg{p}}{\neg{q}}} & {\rulename{Prem}}\\
\ndcorner & \ndhrule  &  {3.}  &  {\con{p}{q}}             & {\rulename{Ass}}\\
\ndvrule  & \ndcorner &  {5.}  &  {\neg{p}}                & {\rulename{Ass}}\\
\ndvrule  & \ndvrule  &  {9.}  &  {p}                      & {3,\rulename{\conop E}}\\
\ndvrule  & \ndvrule  &  {6.}  &  {\falsum}                & {5,9,\rulename{\negop E}}\\
\cline{2-4}
\ndvrule  & \ndcorner &  {7.}  &  {\neg{q}}                & {\rulename{Ass}}\\
\ndvrule  & \ndvrule  &  {10.} &  {q}                      & {3,\rulename{\conop E}}\\
\ndvrule  & \ndvrule  &  {8.}  &  {\falsum}                & {7,10,\rulename{\negop E}}\\
\cline{2-4}
\ndvrule  &           &  {4.}  &  {\falsum}                & {1,5\nto 6,7\nto 8,\rulename{\disop E}}\\
\cline{1-4}
          &           &  {2.}  &  {\neg{\con{p}{q}}}       & {3\nto 4,\rulename{\negop I}}\\
\end{array}$$
First, the premise and conclusion are written as lines 1 and 2, with a generous space between. We then apply \rulename{\negop I} to the conclusion to get a hypothetical deduction with assumption  $\con{p}{q}$ on line 3 and conclusion $\falsum$ on line 4.  Next, we apply \rulename{\disop E} to line 1 to get two nested hypothetical deductions, from $\neg p$ on line 5 to $\falsum$ on line 6, and from  $\neg q$ on line 7 to $\falsum$ on line 8.  The first of these is completed using \rulename{\negop I} to get $p$ on line 9 (justified by \rulename{\conop E} from line 3). The second is completed similarly, with line $q$ on line 10.  In this way, the line numbers match the order of construction of the deduction precisely, which is thereby emphasised to students as they create it. 

The discipline of numbering in the order a deduction is created helps students (and instructors) to think strategically. The goal is to provide a justification for the conclusion given the resources in the premises, and seen this way deduction is just planning how to use the resources to satisfy a goal. While this is a familiar idea in automated reasoning research, it rarely enters the classroom.  By using the above system of numbering, students cannot avoid thinking in this strategic way and learning that introduction rules serve to split the goal into subgoals, whereas elimination rules deploy resources. Strategic concepts such as back-tracking, management of decision points, and an awareness of risk are brought to the fore. Certain rules, such as Disjunction Introduction are seen as ``choice rules" to be used with caution and postponed as long as possible, whereas others, such as 
Implication Introduction are ``automatic" - they can and should be used immediately with no risk of having to undo. 

The use of a new notation has the disadvantage that teaching resources, especially solutions to exercises, have to be produced from scratch. And it was to aid in this that we decided to produce both a {\LaTeX} package for formatting our deductions easily, and a Java app to aid in the generation of both {\LaTeX} code and various other formats for classroom demonstration.

\section{Natural Deduction Planner}
Efficiently creating large numbers of typeset sample deductions can be a daunting prospect. On pen and paper, even a challenging proof can be completed within minutes. However typesetting a proof in software such as {\LaTeX} requires a great deal more effort. With custom packages, structural features such as the scope lines used above can be automated well, but the task of inputting formulas is still cumbersome. Where the hand can draw any symbol at much the speed of any other, typesetting special characters often requires lengthy commands. We began development of a proof assistant software application with the primary goal of overcoming these difficulties, but the result is useful in many more ways than typesetting. We call the result the Natural Deduction Planner (NDP). It generates {\LaTeX} code for use with a custom package. 

Our interface essentially replicates the pen and paper proof process, using the same layout and notation of Gentzen's system, as above. Users input sequents using a set of special characters available onscreen. No special formatting (such as prefix notation) is required - a correctly inputted sequent appears as it would on the page. A range of proof systems are available, such as \rulename{NJ}, \rulename{NK} and Peano Arithmetic. Proofs appear graphically onscreen exactly as they would be typeset. Initially, any premises appear at the top of the window, with a space before the conclusion.

NDP is similar to the Proof Developer tool created by Daniel Velleman.\footnote{Proof Developer is a Java web applet built to accompany Velleman's textbook ``How to Prove It" \cite{velleman1994prove,velleman2006prove} \url{http://www.cs.amherst.edu/~djv/pd/pd.html}} The interfaces are very alike, but where Proof Developer focusses on informal proof writing, NDP is concerned with formal deductions. Both  approaches use ideas of strategy, goals and resources to conduct proofs. Another similar tool is PANDA, developed at the Institut de Recherche en Informatique de Toulouse.\footnote{PANDA is a Java application designed for teaching computer science students in logic, developed by Olivier Gasquet, Fran\c{c}ois Schwarzentruber, and Martin Strecker \url{http://www.irit.fr/panda}} PANDA uses a proof tree style, rather than the Fitch-style calculus implemented in NDP.

At each stage of an NDP deduction, the user can apply any valid rules, and their outcome is immediately displayed.
To apply a rule, a \emph{current goal} must first be selected, which can be any unjustified line. Selecting a current goal allows any possible introduction rules for that line to be applied. Selecting a goal also allows a \emph{current resource} to be selected, enabling its elimination rules. Even if the first step in the proof involves an elimination rule, the user must select a current goal. These explicit realisations of goal and resource help to reinforce the use of strategies in deductions. Rather than beginning with premises working downwards, the user is encouraged to begin at the bottom of the deduction (the goal) and efficiently choose those resources which are needed. By breaking from a strictly linear approach, selecting goals encourages users to consider which available rules are automatic, and which require choice.

\begin{figure}
	\centering
	\includegraphics[width=0.7\linewidth]{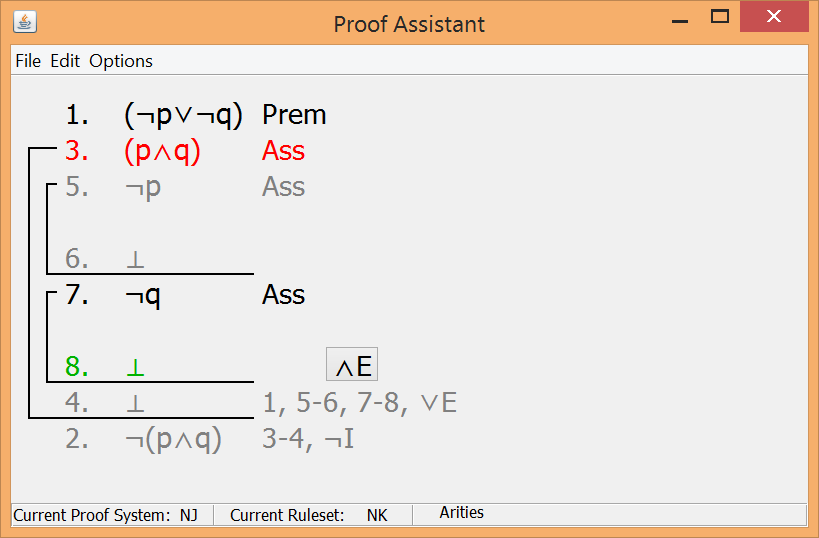}
	\caption{An incomplete proof showing current goal (green), current resource (red) and possible rules (\rulename{\conop E}). All \rulename{NK} rules are available, but currently only \rulename{NJ} rules have been used.}
	\label{fig:screenshot1}
\end{figure}

Lines are numbered in the order of creation, again reinforcing the way the deduction is constructed. This is also useful when reaching a dead-end in the proof - the user can see exactly how they reached this point, and what will happen when they retrace their steps. Highlighting is used to indicate the current goal and resource and also serves to indicate any lines out of scope of the goal. Once all lines have been justified the proof is complete.

By automating the writing of each proof line, users are able to move through deductions faster, focussing more on the strategy involved. Speed and the ability to easily ``undo'' mistakes also removes hurdles from the bulk trial and error method of learning strategy. A student unsure of the next step in a complicated pen and paper proof may be overly wary - a wrong move would result in writing out the whole proof again. On NDP, however, she can chose a rule in the knowledge that the current proof state can easily be retrieved. Similarly, students sometimes find rules like disjunction elimination, which requires creating four new lines and two scopes, to be intimidating and tiresome. Yet disjunction elimination is an automatic rule and should be applied as quickly as possible. In NDP disjunction elimination is achieved with two clicks, a less daunting task.

A \emph{Rule Palette} allows individual rules to be (de)activated independently of the proof system chosen. The rule palette's layout shows the symmetry of Gentzen's rules, and gives some indication as to how the rules fit into different logical systems. By only activating certain rules, students can complete exercises in subsets of a system before being introduced to it fully, and see where certain rules are needed.
For example, the rule palette can be used to demonstrate the importance of double negation elimination in \rulename{NK}, by attempting a proof of $\dis{p}{\neg{p}}$ without double negation elimination to see how far it goes. Once we get stuck,  we turn on double negation elimination to finish the proof. Users can try out their own systems too, to see how different rules interact with each other.

Upon finishing a proof, it can be saved as either an editable proof or a demonstration proof. A demonstration proof has interaction disabled, providing a means to follow through an already complete proof. This is essentially the step by step deduction given above but in electronic form. Editable proofs behave similarly, but allow a user to take over the proof at any point. No work further than completing the deduction is required to generate these.  Proofs can also be exported to  unicode format and as an image. Complete proofs can also easily be animated in \texttt{.gif} format, for use in slides or online.
A primary feature of NDP is its ability to export proofs to {\LaTeX} code. This interacts with a {\LaTeX} package (based on Ti\textit{k}Z) which generates nicely typeset deductions. The task of producing exercises and their solutions involves little more than completing deductions on NDP - no fiddling about with alignment or trying to recall commands required.

We have used NDP as part of a course teaching natural deduction strategies. All the deduction exercises for the course were generated by the software, and it was also made available for students to download. Many students did so, and used NDP to complete exercises and study for tests. We released exercise solutions both as text documents and editable proof files.
A novel use NDP was put to was in catching up on missed lectures. Since NDP applies each rule correctly, by studying what happens students could learn the rule themselves. While the motivation and strategy discussed in lectures was absent here, the correct manipulation of the formula was learned. NDP's automated rule application had some downsides though. Some students found overuse of NDP resulted in over reliance -  you don't have to remember how to set out implication introduction if the software does it for you. Since tests were by pen and paper, this proved problematic. The best combination seemed to be use of pen and paper to practise rules, and NDP to practise strategy.

In the context of tutorials, NDP allowed for greater flexibility in presentation. Again due to ``undo'' it was easier to recover from bad choices, encouraging student participation. Also, a source of potentially confusing transcription errors - the tutor's handwriting - was removed. In one on one situations, NDP allowed for a greater flow of conversation. Discussed strategies for stuck proofs could be implemented quickly and results considered in much less time than would be required to write 10 lines of formulas by hand.

Though originally intended to cover only propositional and predicate logic and Peano Arithmetic, we have begun extending NDP to cover other logics, and to consider new features. We've implemented a system of modal logic and hybrid logic using a labelled deduction method. These rules are available in the standard rule palette but are not thoroughly tested. A very rudimentary second order logic is also available, easily implemented due to Java seeing no distinction between predicate and variable symbols when making substitutions. In an attempt to automate the proof process, a \emph{Magic Mode} is provided. This applies any rules which require no extra input for up to 10 iterations. In exceptional circumstances Magic Mode can complete proofs, but in general will only move forward one or two steps. Finally, a method to include custom axioms has been implemented.

\subsection{Implementation}
NDP was implemented in Java using the \texttt{Swing} and \texttt{SwingX} graphical user interface libraries. The code was written to be extendable and with a goal of modularity, to allow different interfaces to interact with the same backend.

Formulas are held as strings in a {\TeX}  macro format, using prefix notation for easier argument parsing. This also simplifies the process of exporting proofs to {\LaTeX}  code. Each line of a proof is an \texttt{NDLine} object, which contains information such as the formula, the line number and the justification.

A \texttt{ProofMethods} class forms the core of the program. This holds the current proof state as an array of \texttt{NDLine}s.  The application of a rule results in a relevant modification of the proof array and any lines within it. Rules themselves are methods within the \texttt{ProofMethods} class, and the system can be extended by adding new methods to give new rules. In practice, this means that additional rules (such as Disjunctive Syllogism) or extensions to the system can be added fairly easily. \texttt{ProofMethods} is designed to be as self-contained as possible; methods for printing the proof array to the command line mean it could be used without a graphical interface. This is how early development proceeded.

On top of \texttt{ProofMethods} sits the \texttt{ProofPanel} class, a modified \texttt{JPanel} which provides user interaction with \texttt{ProofMethods}. \texttt{ProofPanel} interprets the proof array and arranges the deduction onscreen. The function of the rule palette is implemented entirely within the \texttt{ProofPanel}. If conjunction introduction (\rulename{\conop I}) is turned off then the option to apply that rule becomes unavailable on the \texttt{ProofPanel}. That is, even with \rulename{\conop I} ``disabled'' \texttt{ProofMethods} is still able to apply that rule - there is just no way for the command to do so to reach it. The rule palette makes extensive use of the \texttt{SwingX} library.

A modified \texttt{JFrame} constitutes the main window of the Proof Assistant and controls tasks such as New Proof, Save, Open and Export. Proofs are saved in plain text files which contain complete undo histories and settings profiles. There is no difference between \texttt{.ndp} (editable) and \texttt{.ndu} (demonstration) files - they are read in differently but their contents are identical.

NDP is available on SourceForge at \url{http://sourceforge.net/p/proofassistant/}.


\bibliography{ttl15-pa}
\end{document}